\documentclass{ws-procs10x7}

\begin{document}

\title{\uppercase{CDF and D$\emptyset$ top quark cross section measurements} \newline
\small\it (submitted to ICHEP 2004 conference proceedings)}

\author{L. Sonnenschein}

\address{Boston University, Physics Department, Commonwealth Ave 590, Boston MA 02215,
USA\\E-mail: sonne@bu.edu}

\twocolumn[\maketitle\abstract{
Preliminary results of $t\bar{t}$ cross section measurements and single top 
exclusion limits of the Tevatron experiments CDF and D$\emptyset$ are presented.
The different measurements are based on a dataset between 140 and $200\,\mbox{pb}^{-1}$
corresponding to a data taking period from spring 2002 to September 2003.
The $t\bar{t}$ cross section measurements are consistent with each other and the
Standard Model prediction. Data and Monte-Carlo prediction are in good agreement
in the search for single top quark. Its observation is anticipated with roughly 
$2\,\mbox{fb}^{-1}$ integrated luminosity.}]

\section{Introduction}
The top quark is the heaviest elementary particle discovered yet.
It completes the third generation of quarks as the weak isospin partner
of the bottom quark and has been discovered by the CDF\cite{abe95} and 
D$\emptyset$\cite{aba95}
experiments at the Tevatron $p\bar{p}$ collider during RunI in 1995.
Top quarks are primarily produced in quark antiquark pairs at the Tevatron.
Within the Standard Model the top quark decays almost exclusively into a
$W$ boson and a bottom quark. Conclusively the signature of $t\bar{t}$
production is determined by the decay of the $W$ boson, either leptonically
or hadronically. Decay products of $\tau$ leptons arising from a 
$W$ boson are not distinguished from direct $W$ boson decay particles
of same flavour at this stage of the measurements.

\section{Tevatron Collider}
In comparison to RunI, in which proton antiproton collisions have been taken place
at the Tevatron collider at a center of mass energy of $\sqrt{s}=1.8\,\mbox{TeV}$, 
in RunII the center of mass energy has been elevated 
to $\sqrt{s}=1.96\,\mbox{TeV}$, giving rise to a 30\% increase in the $t\bar{t}$
cross section. The bunch spacing has been reduced from $3500\,\mbox{ns}$ to 
$396\,\mbox{ns}$.
Major upgrades to the Linac and main injector made it possible
to achieve the RunIIa goal of $100 \cdot 10^{30}\mbox{cm}^{-2}\mbox{s}^{-1}$ instantaneous luminosity.
Future improvements are anticipated by a new antiproton recycler and
electron cooling. 

\section{Detector upgrades}
The CDF and D$\emptyset$ detectors have been massively upgraded. Driven by physics 
goals they became very similar. Beside a replaced silicon tracker and 
central drift chamber the 
geometric acceptance of the CDF detector has been increased by new forward 
calorimeters and extended muon coverage. A new silicon tracker, new preshower detectors
and a $2\,\mbox{T}$ superconducting solenoid have been added to the 
D$\emptyset$ detector. Its muon coverage has been extended as well.
The data acquisition and trigger systems have been upgraded for both detectors
to cope with the shorter bunch spacing compared to RunI. 

\section{Data samples}
The RunII physics data taking started in 2002 (February for CDF and July for 
D$\emptyset$). The reported top quark cross section results are 
based on data taken until September 2003 and vary depending on the analysis
channel between 140 and $200\,\mbox{pb}^{-1}$. This has to be compared to about
$110\,\mbox{pb}^{-1}$ accumulated in RunI. 
Over $400\,\mbox{pb}^{-1}$ of data are already recorded in RunII. 
Analysis of subsequent data is in progress.

\section{Top quark antiquark pair production cross section measurements}
The Standard Model prediction for the $t\bar{t}$ production cross section
at Tevatron in RunII with a center of mass energy of $1.96\,\mbox{TeV}$ and a
given top quark mass of $m_t=175\,\mbox{GeV}$ amounts in Next-to-Leading order
calculation\cite{ca04} \cite{ki04} to 
\begin{eqnarray}
  \sigma_{\mbox{\scriptsize NLO}}(t\bar{t})=6.7^{+0.7}_{-0.9}\,\mbox{pb} \nonumber
\end{eqnarray}
taking into account uncertainties in the proton probability
density functions.

\begin{figure*}
\epsfxsize30pc
\unitlength 1cm
\begin{picture}(6.,7.7)
  \put(-0.6,-1.0){\includegraphics[width=7.0cm]{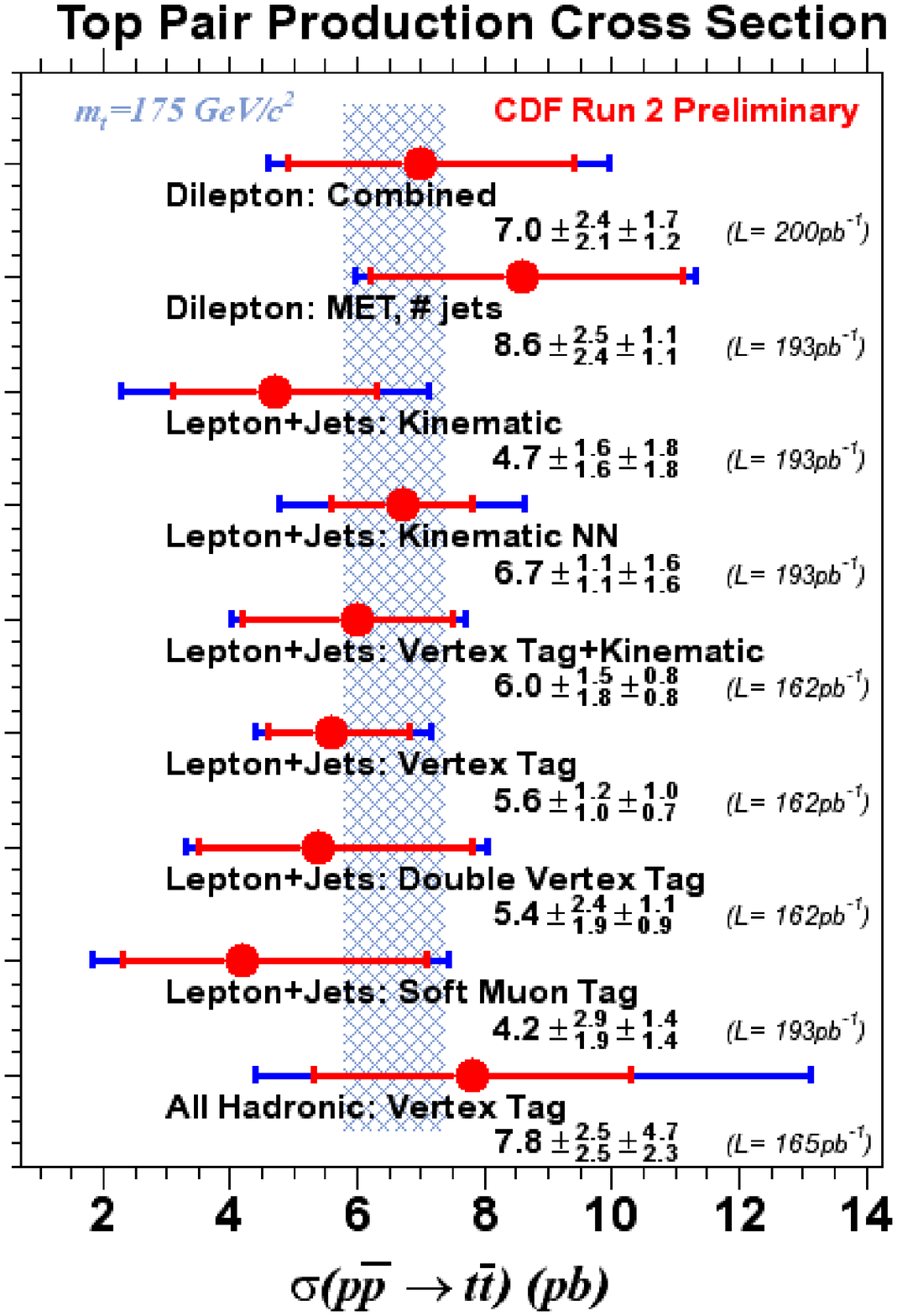}}
  \put(6.4,-0.1){\includegraphics*[angle=0, width=8.7cm]{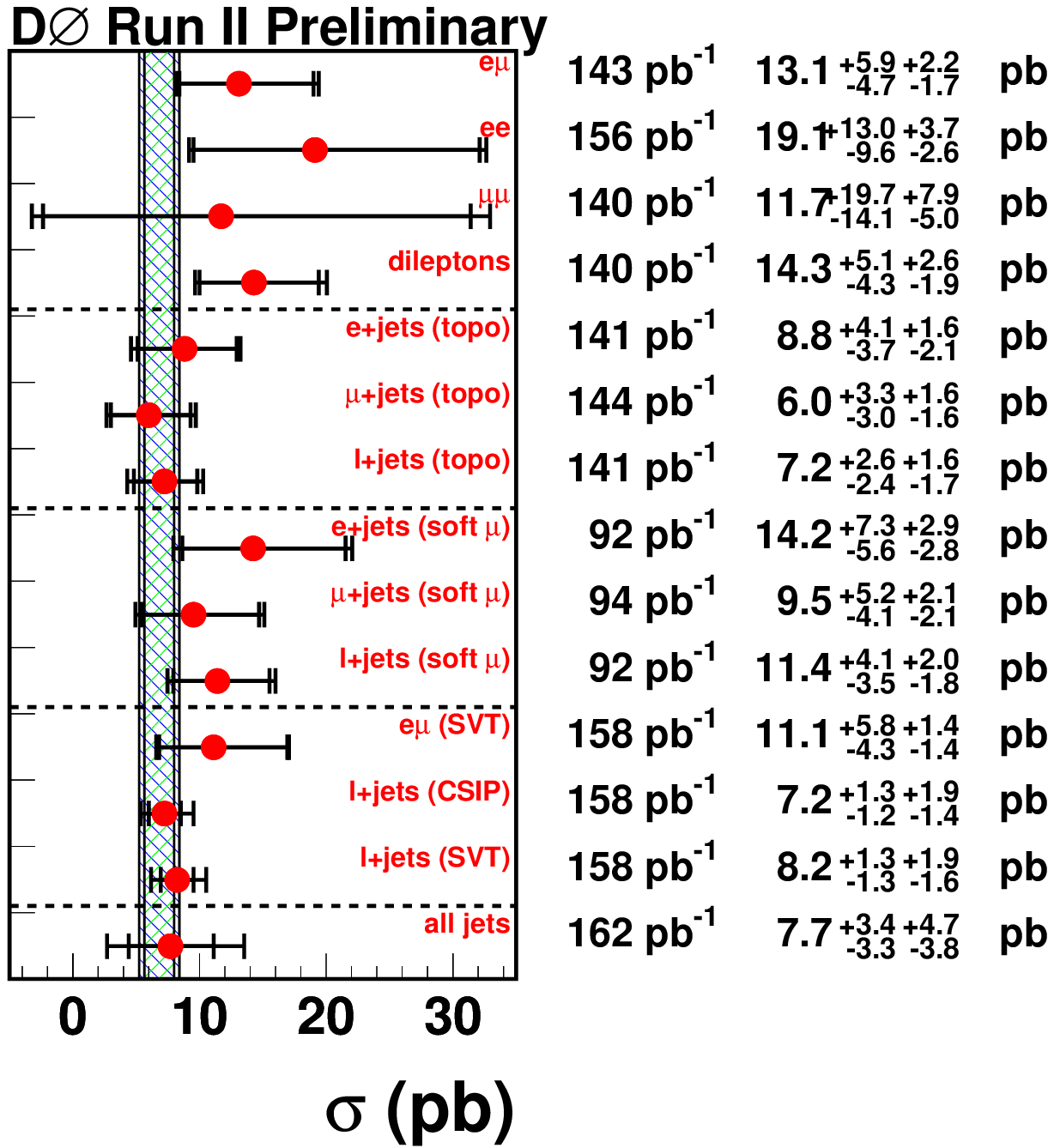}} 
\end{picture}
\caption{Top quark antiquark pair cross section summary of the CDF and D$\emptyset$
experiments. For comparison the Standard Model Next-to-Leading order prediction 
for a top quark mass of $m_t=175\,\mbox{GeV}$ including uncertainties on the top 
quark mass (CDF + D$\emptyset$) and proton probability density functions (D$\emptyset$) 
is superposed.
\label{ttbarxsecsummary}}
\end{figure*}

\subsection{Dilepton channel}
The dilepton channel can be characterized by a small branching ratio
and relatively few Standard Model backgrounds which are Drell-Yan 
(in case of like-flavoured leptons) and diboson production, 
missing transverse energy and isolated lepton fakes.
As selection criteria at least two high transverse momentum jets, large missing energy
to account for the neutrinos and two oppositely charged leptons are required.
CDF requires optionally the second lepton to be very loose, even just an isolated
high transverse momentum track. The cross sections derived by CDF with an integrated luminosity of about $200\,\mbox{pb}^{-1}$ and D$\emptyset$ with $140\,\mbox{pb}^{-1}$ are
\begin{eqnarray}
\sigma_{\mbox{\tiny CDF}} & = & 7.0^{+2.4}_{-2.1}(\mbox{stat})^{+1.6}_{-1.1}(\mbox{sys})\pm0.4(\mbox{lum})\,\mbox{pb}  \nonumber \\ 
\sigma_{\mbox{\tiny D$\emptyset$}} & = & 14.3^{+5.1}_{-4.3}(\mbox{stat})^{+2.6}_{-1.9}(\mbox{sys})\pm0.9(\mbox{lum})\,\mbox{pb} . \nonumber 
\end{eqnarray} 

CDF pursues also an alternative approach to counting experiments. 
Merely two leptons are 
required and significant missing transverse energy in case of like-flavoured leptons.
Data are fitted with $t\bar{t}$, $WW$ and $Z\rightarrow \tau\tau$ templates 
in a two dimensional phase space of missing transverse energy and
number of jets. The $t\bar{t}$ signal and $WW$ diboson background cross sections are
determined with about $200\,\mbox{pb}^{-1}$ to be
\begin{eqnarray}
\sigma_{\mbox{\tiny CDF}}(t\bar{t}) & = & 8.6^{+2.5}_{-2.4}(\mbox{stat})\pm1.1(\mbox{sys})\,\mbox{pb} \nonumber \\ 
\sigma_{\mbox{\tiny CDF}}(WW) & = & 12.6^{+3.2}_{-3.0}(\mbox{stat})\pm1.2(\mbox{sys})\,\mbox{pb} \nonumber 
\end{eqnarray} 
(luminosity uncertainties are omitted here and from now on where negligible).

In the $e\mu$-channel of the $t\bar{t}$ production D$\emptyset$ exploits the powerful background rejection of a secondary vertex tagger (SVT) requiring at least one $b$-tagged jet after preselection cuts. A fit to
the jet multiplicity distribution yields
\begin{eqnarray}
\sigma_{\mbox{\tiny D$\emptyset$}}=11.1^{+5.8}_{-4.3}(\mbox{stat})\pm1.4(\mbox{sys})\pm0.7(\mbox{lum})\,\mbox{pb} \nonumber
\end{eqnarray} 
with an integrated luminosity of about $160\,\mbox{pb}^{-1}$.

\subsection{Lepton plus jets channel}
The lepton plus jets channel can be distinguished by a large branching ratio,
$W$ boson plus jets background and multijet background with one jet faking a lepton
and missing transverse energy arising from mismeasured jet energies.
Preselection cuts are demanding one charged high transverse momentum lepton,
multiple high transverse momentum jets and large missing transverse energy.

In the topological analyses, variables with maximal significance and 
minimal jet energy scale dependence - since dominating systematics - 
are chosen. Variables like the transverse energy of the event $H_T$ are fitted directly.
Alternatively they are combined into a likelihood discriminant either by hand (Discr.) 
or via a neural network (NN).
The measured cross sections for the $e$ and $\mu$ lepton plus jet channels combined are
\begin{eqnarray}
\sigma_{\mbox{\tiny CDF}}^{H_T} & = & 6.7\pm1.1(\mbox{stat})\pm1.6(\mbox{sys})\,\mbox{pb}  \nonumber \\ 
\sigma_{\mbox{\tiny CDF}}^{NN} & = & 4.7\pm1.6(\mbox{stat})\pm1.8(\mbox{sys})\,\mbox{pb}  \nonumber \\ 
\sigma_{\mbox{\tiny D$\emptyset$}}^{\mbox{\tiny Discr.}} & = & 7.2^{+2.6}_{-2.4}(\mbox{stat})^{+1.6}_{-1.7}(\mbox{sys})\pm0.5(\mbox{lum})\,\mbox{pb} \nonumber 
\end{eqnarray} 
with an integrated luminosity of about $195\,\mbox{pb}^{-1}$ and $140\,\mbox{pb}^{-1}$
for the CDF and D$\emptyset$ experiments respectively.

Several CDF analyses in the lepton plus jets channel require at least
one jet to be $b$-tagged either with a secondary vertex or a soft lepton tagger
which exploits low momentum leptons within jets.
The cross section is obtained by accumulation of the heavy flavour fraction (HF)
and data normalization to the jet multiplicity distribution.
Another way to obtain the cross section is to apply a kinematic fit to the shape
of the leading jet transverse energy ($E_{\perp\max}^{\mbox{\scriptsize jet}}$)
distribution. Both cross section measurements take advantage of about 
$160\,\mbox{pb}^{-1}$ integrated luminosity.
In contrast to the first two methods which make use of a secondary 
vertex tagger a soft lepton tagger (SLT) can be applied. Here the measurement  
has been accomplished by a counting experiment in the muon plus jets channel
with an integrated luminosity of about $195\,\mbox{pb}^{-1}$.
The measured cross sections are given by  
\begin{eqnarray}
\sigma_{\mbox{\tiny CDF}}^{SVT, HF} & = & 5.6^{+1.2}_{-1.0}(\mbox{stat})^{+1.0}_{-0.7}(\mbox{sys})\,\mbox{pb}  \nonumber \\ 
\sigma_{\mbox{\tiny CDF}}^{SVT, E_{\perp\max}^{\mbox{\tiny jet}}} & = & 6.0^{+1.5}_{-1.8}(\mbox{stat})\pm0.8(\mbox{sys})\,\mbox{pb}  \nonumber \\[1ex] 
\sigma_{\mbox{\tiny CDF}}^{SLT, \ell=\mu} & = & 4.2^{+2.9}_{-1.9}(\mbox{stat})\pm1.4(\mbox{sys})\,\mbox{pb} . \nonumber 
\end{eqnarray} 
Yet another variation pursued by CDF is to require two jets to be $b$-tagged with the 
secondary vertex tagger. This analysis has been subdivided into a standard  
($\epsilon_b^{\max}\simeq0.4$) and a higher
efficiency $b$-tagging version ($\epsilon_b^{\max}\simeq0.5$) 
whose primary purpose is not the cross section 
measurement but to reduce combinatorics in future top quark mass measurements.
The cross section of both variants has been determined to be
\begin{eqnarray}
\sigma_{\mbox{\tiny CDF}}^{SVT, \mbox{\tiny Standard}} & = & 5.4^{+2.4}_{-1.9}(\mbox{stat})^{+1.1}_{-0.9}(\mbox{sys})\,\mbox{pb}  \nonumber \\ 
\sigma_{\mbox{\tiny CDF}}^{SVT, \mbox{\tiny High eff.}} & = & 8.2^{+2.4}_{-2.1}(\mbox{stat})^{+1.8}_{-1.0}(\mbox{sys})\,\mbox{pb} \; .  \nonumber 
\end{eqnarray}

\begin{table*}
\begin{center}
\begin{tabular}{|l||c|c|c|c|} \hline
 Cross sections & $s$-channel & $t$-channel & $s+t$-channel & $\cal{L}$ \\ \hline \hline
 SM prediction (NLO) & $\sigma=0.88\,\mbox{pb}$ & $\sigma=1.98\,\mbox{pb}$ & 
 $\sigma=2.86\,\mbox{pb}$ & --- \\ \hline
 \raisebox{0.3ex}[-0.3ex]{CDF (@ 95\% C.L.)} & \raisebox{0.3ex}[-0.3ex]{$\sigma<13.6\,\mbox{pb}$} & \raisebox{0.3ex}[-0.3ex]{$\sigma<10.1\,\mbox{pb}$} & 
\raisebox{0.3ex}[-0.3ex]{$\sigma<17.8\,\mbox{pb}$} & 
\rule{0ex}{3.0ex} \raisebox{0.3ex}[-0.3ex]{$160\,\mbox{pb}^{-1}$} \\ \hline
 \raisebox{0.3ex}[-0.3ex]{D$\emptyset$ (@ 95\% C.L.)} & \raisebox{0.3ex}[-0.3ex]{$\sigma<19\,\mbox{pb}$} & \raisebox{0.3ex}[-0.3ex]{$\sigma<25\,\mbox{pb}$} & 
\raisebox{0.3ex}[-0.3ex]{$\sigma<23\,\mbox{pb}$} & \rule{0ex}{3.0ex} \raisebox{0.3ex}[-0.3ex]{$156-169\,\mbox{pb}^{-1}$} \\ \hline
\end{tabular}
\end{center}
\caption{Single top quark cross section predictions for a top quark mass of $m_t=175\,\mbox{GeV}$ and measurement limits of the CDF and D$\emptyset$ experiments at the 95\% confidence limit.}\label{singletopxsec}
\end{table*}

\subsection{All hadronic channel}
The largest branching fraction accompanied by the largest background 
is inherent to the all hadronic channel. To reduce the multijet background
which exceeds the signal by three orders of magnitude topological cuts and
$b$-tagging has to be applied (here SVT single tagged events have been required). 
In addition vetos against isolated leptons and poorly reconstructed primary 
vertices help to enhance the significance. 
While CDF applies cuts onto topological variables in one or two dimensions 
to count the excess of single tagged events in the 
$6<N_{\mbox{\scriptsize jets}}<8$ signal region of the jet multiplicity distribution
D$\emptyset$ applies a cascade of three neural networks onto topological variables
to enhance the significance by cutting on a final discriminant before counting 
event excess.
The cross section measurements yield
\begin{eqnarray}
\sigma_{\mbox{\tiny CDF}} & = & 7.8^{+2.5}_{-1.0}(\mbox{stat})^{+4.7}_{-2.3}(\mbox{sys})\,\mbox{pb}  \nonumber \\ 
\sigma_{\mbox{\tiny D$\emptyset$}} & = & 7.7^{+3.4}_{-3.3}(\mbox{stat})^{+4.7}_{-3.8}(\mbox{sys})\pm0.5(\mbox{lum})\,\mbox{pb}  \nonumber 
\end{eqnarray} 
for an integrated luminosity of about $165\,\mbox{pb}^{-1}$ for CDF and about 
$160\,\mbox{pb}^{-1}$ for D$\emptyset$.

The summary of all CDF and D$\emptyset$ $t\bar{t}$ cross section measurements is given
in fig. \ref{ttbarxsecsummary}.

\section{Single top quark search}
The Standard Model prediction for single top quark production at 
Tevatron in RunII with a center of mass energy of $1.96\,\mbox{TeV}$ and a
given top quark mass of $m_t=175\,\mbox{GeV}$ amounts in Next-to-Leading order
calculation for the $t$-channel\cite{st97} \cite{ha02} and 
$s$-channel\cite{sm96} respectively to
\begin{eqnarray}
  \sigma_{\mbox{\scriptsize NLO}}^t & = & 1.98\pm0.21\,\mbox{pb} \nonumber \\
  \sigma_{\mbox{\scriptsize NLO}}^s & = & 0.88\pm0.07\,\mbox{pb} \; . \nonumber 
\end{eqnarray}
Single top quarks are produced by electroweak interactions. $s$-channel 
cross section measurements can be exploited to obtain limits on the 
magnitude of the CKM matrix element $V_{tb}$.

Single top quark events are selected in requiring one isolated high 
transverse momentum lepton, large missing transverse energy and two high transverse 
momentum jets of which at least one has to be $b$-tagged. 
$W$-boson plus jets and misidentified $t\bar{t}$ events constitute the dominating
background. Data and Monte-Carlo predictions are in good agreement.
Cross section limits at the 95\% confidence level for the $s$, $t$ and $s+t$-channels
combined are given in table \ref{singletopxsec}.
Observation of single top quark production is expected with an integrated luminosity
of about $2\,\mbox{fb}^{-1}$.

\section*{Acknowledgments}
Many thanks to the stuff members at Fermilab, collaborating institutions
and the top physics group convenors of the CDF and D$\emptyset$
experiments for their support and help. In particular Evelyn Thomson merits 
special thanks for her tireless support and cooperativeness in many 
productive discussions.


\end{document}